\documentclass[12pt,preprint]{aastex}
\usepackage{lineno}
\usepackage{amsmath}
\usepackage{color}
\RequirePackage{lineno}

\def\pcs{photons cm$^{-2}$ s$^{-1}$}
\def\pcssp{photons cm$^{-2}$ s$^{-1}$ }
\def\psrsp{PSR J0007+7303 }
\def\psr{PSR J0007+7303}
\def\ltsima{$\; \buildrel < \over \sim \;$}
\def\simlt{\lower.5ex\hbox{\ltsima}} 
\def\gtsima{$\; \buildrel > \over \sim \;$}
\def\simgt{\lower.5ex\hbox{\gtsima}} 
\def\arcsec{\hbox{$^{\prime\prime}$}}
\def\deg{\hbox{$^\circ$}}
\def\degsp{\hbox{$^\circ$ }}
\def\f{\textit{Fermi}}
\def\fermis{\textit{Fermi }}
\def\g{\hbox{$\gamma$}}

\def\lsp{LAT }
\def\c{CTA 1 }
\def\s{SNR }
\def\x{\textit{XMM-Newton }}
\def\C{\textit{Chandra }}
\def\r{RX J0007.0+7302 }

\begin{document}
\title{PSR J0007+7303 in the CTA1 SNR: New Gamma-ray Results from Two Years of
  \f-LAT Observations} 
\author{
A.~A.~Abdo\altaffilmark{1},
K.~S.~Wood\altaffilmark{2},  
M.~E.~DeCesar\altaffilmark{3,4}, 
F.~Gargano\altaffilmark{5}, 
F.~Giordano\altaffilmark{5,6},
P.~S.~Ray\altaffilmark{2},
D.~Parent\altaffilmark{1},  
A.~K.~Harding\altaffilmark{3}, 
M.~Coleman~Miller\altaffilmark{7},
D.~L.~Wood\altaffilmark{8}, 
M.~T.~Wolff\altaffilmark{2}
}
\altaffiltext{1}{Center for Earth Observing and Space Research, College of Science, George Mason University, Fairfax, VA 22030, resident at Naval Research Laboratory, Washington, DC 20375}
\altaffiltext{2}{Space Science Division, Naval Research Laboratory, Washington, DC 20375-5352}
\altaffiltext{3}{NASA Goddard Space Flight Center, Greenbelt, MD 20771, USA}
\altaffiltext{4}{Department of Physics and Department of Astronomy, University of Maryland, College Park, MD 20742}
\altaffiltext{5}{Istituto Nazionale di Fisica Nucleare, Sezione di Bari, 70126 Bari, Italy}
\altaffiltext{6}{Dipartimento di Fisica ``M. Merlin" dell'Universit\`a e del Politecnico di Bari, I-70126 Bari, Italy}
\altaffiltext{7}{Department of Astronomy, University of Maryland, College Park, MD 20742}
\altaffiltext{8}{Praxis Inc., Alexandria, VA 22303, resident at Naval Research Laboratory, Washington, DC 20375}
\begin{abstract}

\end{abstract}
One of the main results of the \textit{Fermi Gamma-Ray Space
  Telescope} is the discovery of \g-ray selected pulsars. The high
magnetic field pulsar, PSR J0007+7303 in CTA1, was the first ever to
be discovered through its \g-ray pulsations. Based on analysis of 2
years of LAT survey data, we report on the discovery of \g-ray
emission in the off-pulse phase interval at the $\sim 6 \sigma$
level. The flux from this emission in the energy range E $\geq$ 100
MeV is $F_{100} = (1.73 \pm 0.40) \times 10^{-8}$ \pcssp and is best
fitted by a power law with a photon index of $\Gamma = 2.54 \pm 0.14$.
The pulsed \g-ray flux in the same energy range is $F_{100} = (3.95
\pm 0.07) \times 10^{-7} $ \pcssp and is best fitted by an
exponentially-cutoff power-law spectrum with a photon index of $
\Gamma = 1.41 \pm 0.23$ and a cutoff energy $E_c = 4.04 \pm 0.20$ GeV.
We find no flux variability neither at the 2009 May glitch nor in the
long term behavior. We model the \g-ray light curve with two
high-altitude emission models, the outer gap and slot gap, and find
that the model that best fits the data depends strongly on the assumed
origin of the off-pulse emission.  Both models favor a large angle
between the magnetic axis and observer line of sight, consistent with
the nondetection of radio emission being a geometrical effect. Finally
we discuss how the LAT results bear on the understanding of the
cooling of this neutron star.


\section{Introduction\label{section-intro}}
The \c supernova remnant (SNR) (G119.5+10.2) is a composite \s
characterized by a large radio shell enclosing a smaller pulsar wind
nebula (PWN). 
\cite{Pineault1993} derived a kinematic distance of $1.4\pm0.3$ kpc
based on associating an \ion{H}{1} shell found to the northwestern
part of the remnant to the remnant itself. Observations in X-rays with
\textit{ASCA} and \textit{ROSAT} revealed a central filled SNR with
emission extending to the radio shell in the south and southeast as
well as in the north and northwest radio-quiet regions
\citep{Seward1995}. X-ray observations with {\em ROSAT} also revealed
the point source RX J0007.0+7302 \citep{Seward1995}. X-ray
observations with \C revealed a compact PWN and a jet-like structure
\citep{Halpern2004}. Radio and X-ray characteristics of \c imply an
age in the range 5000--15000 years
\citep{Pineault1993,Slane1997,Slane2004}. \cite{Slane2004} estimated
for the age of the SNR a value of $1.3 \times 10^4 \, d_{1.4} $ yr
(where $d_{1.4}$ is the distance in units of 1.4 kpc) which is in good
agreement with the spin-down age estimate of 14,000 yr from \g-rays
\citep{CTA1}.  The offset of the point source \r from the geometrical
center of the radio SNR allows for the estimate of the transverse
velocity of the point source which \cite{Slane2004} estimated to be
$\sim$450 km s$^{-1}$.

Prior to the launch of \fermis the EGRET \g-ray source 3EG J0010+7309,
which lies within the boundaries of the radio SNR, showed the
characteristics of a pulsar
\citep{Brazier1998,Halpern2004}. \cite{mattox1996_pulsar} discussed
this source as a potential candidate for a radio-quiet \g-ray
pulsar. A search for \g-ray pulsations using EGRET data didn't reveal
the pulsar \citep{Ziegler2008}. The characteristics of this source in
X-rays also pointed to it as a pulsar \citep{Halpern2004}. In fact
\cite{Halpern2004} called the source the ``pulsar'' although no
pulsations were detected from this source using FFT searches on
$\sim26$ ks of \x data \citep{Slane2004} or in radio with the Green
Bank Telescope \citep{Halpern2004}. Very deep searches for a
counterpart for \r in optical and radio resulted only in upper limits
\citep{Halpern2004}. By correlating X-ray images from \C with those in
the optical waveband of the \c region, \citet{Halpern2004} gave the
most accurate position of this source to date ((J2000.0)
00$^\mathrm{h}$07$^\mathrm{m}$1\fs56,+73\degr03\arcmin08.1\arcsec)
with an accuracy of $\sim 0\arcsec.1$. The discovery of the pulsar
didn't come until the launch of \f. During its commissioning stage the
\lsp discovered a 315.87 ms pulsar at the location of RX J0007+7303
\citep{CTA1}. The discovery of the pulsar in \g-rays prompted a long,
$\sim130$ ks, \x observation to search for pulsations from this source
(PI: Caraveo 2008, ObsID: 06049401). Using this \x data set along with
the timing model from the \lsp \citep{BSP} X-ray pulsations from this
source in X-rays were finally detected \citep{Lin2010,Caraveo2010}.

This paper reports further analysis on the LAT data and related
observations of the \c SNR and its \g-ray pulsar, extending the
initial report of the pulsar discovery \citep{CTA1}.  The new
developments contribute to a characterization both of the pulsar and its
relation to the associated extended source, exploiting two years of
data now accumulated by \f.  New developments fall in three
distinct areas.  First, timing analysis of the pulsar spanning the two
years shows a glitch near the middle of the interval at MJD 54952.652
(1 May 2009), with relative change in pulse frequency $\Delta\nu/\nu \sim 6 \times
10^{-7}$.  This permits careful comparison of the pulsar
characteristics before and after the glitch.  Second, we present new
\g-ray results concerning the relationship of the pulsar to the
surrounding remnant, the first detection of an extended source in the
off-pulse emission. Finally, we summarize the multi-wavelength picture
of the source. From this perspective the outstanding characteristic of
the neutron star in CTA1 is that it appears cool for its inferred age.
We reconsider both the temperature and the age estimates, and then
relate this to pulsar characteristics established from timing LAT
\g-rays.  In earlier work \citep{Halpern2004} the pulsar in \c has been
assigned an inferred mass exceeding 1.42 $M_\sun$ even though it is
not in a binary, because the larger mass should accelerate cooling.
Even from initial timing solutions it was clear that this pulsar has a
very strong magnetic field, though weaker than that of a magnetar,
hence it is atypical in two respects.


\section{Gamma-Ray Observations and Data Analysis}
We used 2 years of survey data collected with the \lsp to study this
source in \g-rays. The data set starts 2008 August 4 and ends 2010
August 4 (54682.68 - 55412.65 MJD). This large data sample with high
statistics, compared to the six weeks of observations used for the
discovery paper \citep{CTA1} and the six months used for the \fermis
pulsar catalog paper \citep{psrcat} allows us to perform several key
timing and spectral analyses not feasible in prior studies. In
particular we study the phase-resolved spectra and the flux
variability especially around the 2009 May 1 glitch, we search for
off-pulse emission and build a precise timing model. Throughout this
paper we used ``diffuse'' class photons events with the P6$\_$V11
instrument response functions
(IRFs)\footnote{http://fermi.gsfc.nasa.gov/ssc/data/analysis/documentation/Cicerone/Cicerone\_LAT\_IRFs/IRF\_overview.html}
\citep{LATinstrumentshort}. To reject atmospheric \g-rays from the
Earth's limb, we selected events with zenith angle $<100\deg$.

\subsection{Timing Analysis}
\label{sec:timing}

For the pulse timing analysis, we selected events with energies $>
170$ MeV that were reconstructed within 1.3\degsp from the
\textit{Chandra} location for RX J0007.0+7303 source
\citep{Halpern2004}.  These radius and energy cuts were selected to
maximize the pulsed significance. Following the procedure described by
\citet{BlindTiming}, we measured a total of 72 pulse times of arrival
(TOAs), each with a integration time of about 10 days, referenced to
the geocenter.  Each TOA was determined using the unbinned maximum
likelihood technique from \citet{BlindTiming} using a template profile
consisting of 2 gaussian components.  The TOAs were then fit to a
timing model using \textsc{Tempo2} \citep{Hobbs2006}.  The model,
shown in Table \ref{table:timing}, includes position, frequency
($\nu$) and the first three frequency derivatives
($\dot{\nu},\ddot{\nu}, \dddot{\nu}$). We fitted for position since
this will give the smallest timing residuals and will allow for the
comparison of our timing position to that from {\em Chandra}. We find
our timing position, shown in Table \ref{table:timing}, to be
consistent with the {\em Chandra} position of compact source RX
J0007.0+7303 (00:07:01.56, 73:03:08.3; see \cite{Halpern2004}).  In
addition, a glitch on 2009 May 1 (MJD 54952.652) was included with an
instantaneous step in $\nu$ and $\dot\nu$ at the glitch. The glitch
epoch was chosen to produce a zero phase jump at the glitch.  For this
glitch we measure $\Delta \nu/\nu =5.54(1) \times 10^{-7}$.  With a
shorter dataset, \citet{BlindTiming} could not be certain of the step
in $\dot{\nu}$ at the glitch. Our extended observations allow us to be
confident of the $\Delta\dot{\nu}/\dot{\nu} = 9.7(6) \times 10^{-4}$
at the glitch.

The timing model includes second and third frequency derivatives,
which are required to obtain white residuals. This is presumably an
indication of timing noise in this young pulsar.  We note that for a
braking index of 3, a $\ddot{\nu}$ of $1.24\times10^{-23} $ s$^{-3}$
is expected from the secular spin down of the pulsar.  This accounts
for about a third of the total measured $\ddot{\nu}$.  If one
interprets the measured $\ddot{\nu}$ as being entirely due to the
secular spin down of the pulsar, one obtains a braking index ($n$) of
$n = \nu\ddot{\nu}/\dot{\nu}^2 = 9.95$.

\subsection{Detection of Off-Pulse Emission}
\label{sec:offpulse}
\begin{figure}
\begin{center}
\includegraphics[scale=0.8,angle=0]{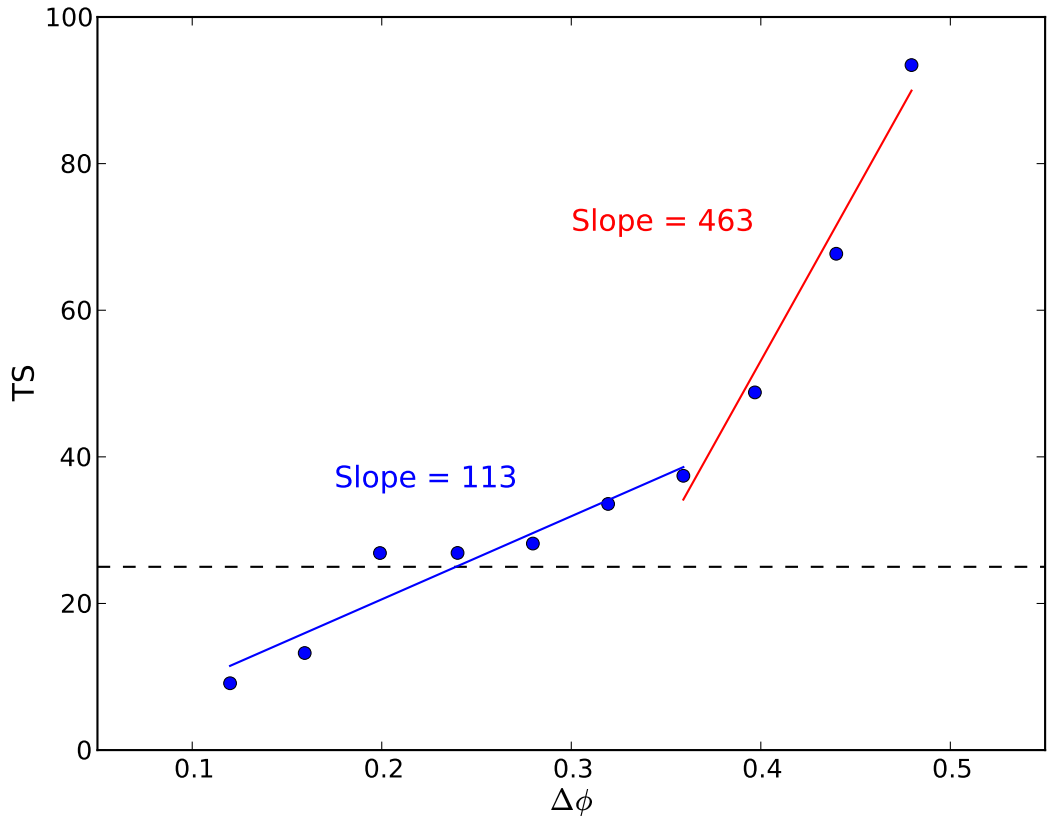}
\caption{Test Statistic trend versus phase-bin widths. The dashed line
  represent the detection threshold (TS = 25). In all of the bins the
  center value was $\phi = 0.89$.}
\label{fig:TS_bins}
\end{center}
\end{figure}

To search for any \g-ray emission present in the off-pulse part of the
phase, we had first to determine accurately the definition of the
off-pulse phase window. One might simply determine the off-pulse
interval by eye. In that method the off-pulse starts when any apparent
pulsed emission decreases to the levels of the background and ends
when the pulsed emission resumes and an increase above background is
seen. A less arbitrary method is to perform a likelihood analysis in
small phase bins, gradually increasing the width of consecutive
bins. For all of these width-varying bins only the left and right
limits are changing while the center of the bin is fixed at what one
believes to be the center of the off-pulse interval. A spectral shape
is assumed for any off-pulse emission and the increase in signal as a
function of phase bin width ($\Delta\phi$) is
analysed.\\
We performed this analysis with \texttt{gtlike} where we selected a
region of interest of 20$\degr$ and a source region of $30\degr$ (see
\S \ref{sec:spectral} for more details). In this analysis all phase
bins were centered at $\phi =0.89$. In the case of absence of signal
in off-pulse, one would expect the distribution of the test statistic
(TS) values \citep{mattox1996} to be centered around zero with no
correlation with the width of the phase bins $\Delta \phi$.
However, in the case of
presence of off-pulse emission one would expect the significance to
increase linearly with the increase in the width of the phase bins
until one starts integrating photons from the pulsar itself where a
sharp increase in significance is then seen. The off-pulse width is
then defined as the point at which the sharp increase in the TS
occurs. This is shown in Figure \ref{fig:TS_bins}.

From the plot, one sees two ranges of data points, with a clear break
in slope between the earlier range (first seven points) and the
remainder. We fitted first order polynomials to each range and
detected a clear significant change in slope, by about a factor of
four. This is taken to be the onset of contamination from the pulsed signal. We
therefore define the off-pulse interval to be $ \phi \in [0.71 -1.07]$
with a width of $\Delta \phi=0.36$. \\

As can be seen from Figure \ref{fig:TS_bins} there is a significant
detection of \g-ray emission in the off-pulse phase. At a TS of $\sim
40$ this is the first detection of \g-ray emission in the off-pulse of
PSR J0007+7303. An earlier \f-\lsp survey of PWNe by \cite{LATPWNCat},
using 16 months of \lsp data, noted a candidate for an off-pulse
emission from \psrsp but it was below the detection threshold even
though a wider phase window was used for the off-pulse. The present
detection, using 2 years of \lsp data but with a conservative phase
window is unambiguous.

\subsubsection{Off-Pulse Extension Analysis}
\begin{figure}
\begin{center}
\includegraphics[scale=0.7,angle=0]{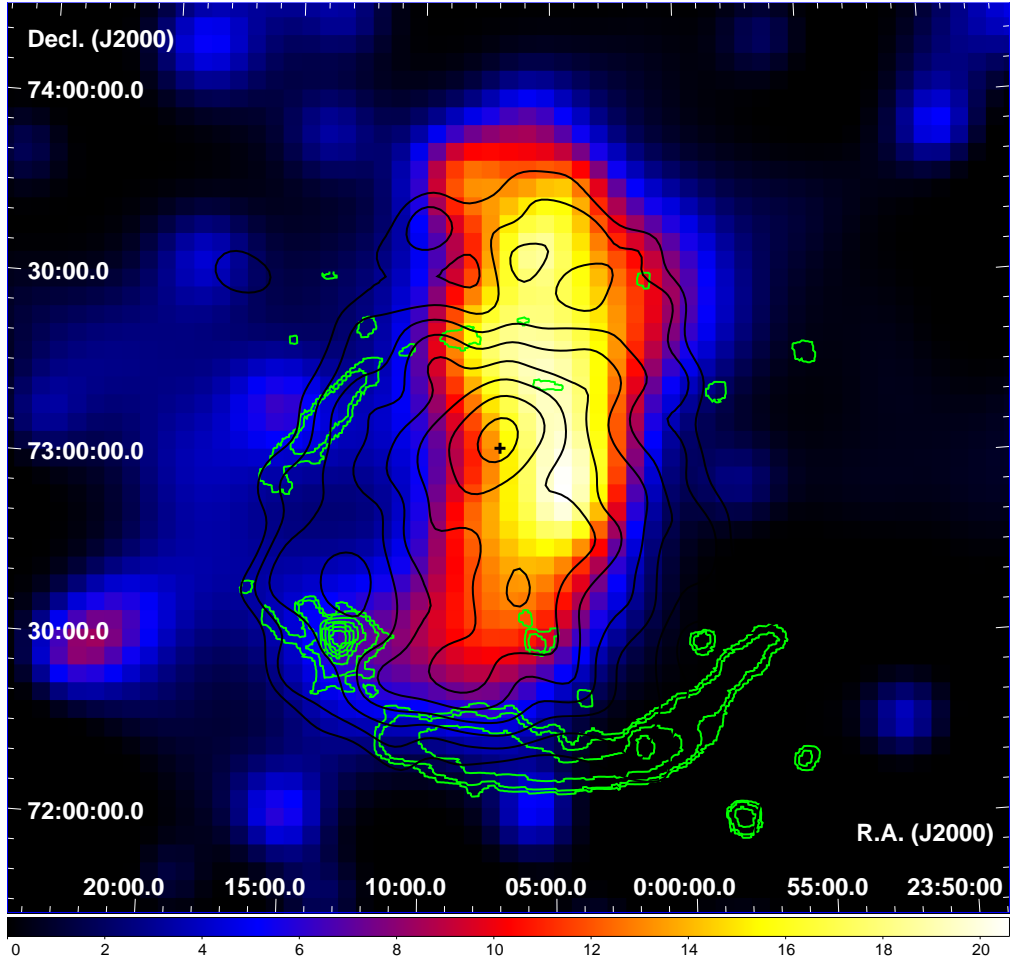}
\caption{\f-\lsp TS map of the off-pulse part of \psrsp ($ \phi \in
  [0.71 -1.07]$).  Black cross marks the location of the pulsar.
  \textit{ROSAT} X-ray contours are shown in black \citep{Seward1995}. Radio
  contours are shown in green \citep{Pineault1997}.}
\label{fig:TS_map}
\end{center}
\end{figure}

Figure \ref{fig:TS_map} shows a TS map of the off-pulse part of
\psr. On the same Figure we show \textit{ROSAT} X-ray contours in black
\citep{Seward1995} and radio contours in green \citep{Pineault1997}.
From the Figure one can see that 1) the \g-ray signal detected with
the \lsp is better correlated with the \textit{ROSAT}  X-ray emission than with
the radio SNR and 2) that the off-pulse \g-ray emission is clearly
extended.

To check for a possible extension in the off-pulse \g-ray emission, we
performed the likelihood analysis similar to that in \S
\ref{sec:offpulsespec} with the further addition of an extended disk
template to describe the possible extension of the emission. Different
angular sizes in the range 0.1$\deg$--1.0$\degsp$ and centroid
position templates have been fitted. In each case the significance was
compared to a simple point-like hypothesis. A disk of radius
$0.6\deg\pm0.3\degsp $ is favored, and a point-like hypothesis is
excluded at the 95\% confidence level. A template in the shape of an
ellipse was also fitted to the off-pulse emission. We found no
compelling statistical evidence in favor of this fit compared to the
disk.

\subsection{Spectral Analysis}
\label{sec:spectral}
Spectral analyses for this source were performed using the \f\ LAT
maximum-likelihood Science Tool \texttt{gtlike} in its binned
mode\footnote{http://fermi.gsfc.nasa.gov/ssc/data/analysis/}. Fits
were performed on a $14\deg \times 14 \deg$ region of the sky centered
at the pulsar position selecting photons in the energy range 0.1 to
300 GeV.  We used a model that included diffuse emission components as
well as nearby \g-ray sources from the First \f-\lsp \g-ray catalog
(1FGL) \citep{Fermicatalog} that fell within 19\degsp from the
position of \psrsp. The Galactic diffuse emission was modeled using
the \texttt{gll$\_$iem$\_$v02$\_$P6$\_$V11$\_$DIFFUSE} model and the
isotropic background using the
\texttt{isotropic$\_$iem$\_$v02$\_$P6$\_$V11$\_$DIFFUSE}
model\footnote{http://fermi.gsfc.nasa.gov/ssc/data/access/lat/BackgroundModels.html}.

In performing the fit we fixed all the parameters of the sources that
fell between 14$^\circ$ and 19$^\circ$ from PSR J0007+7303 to their
values in the 1FGL catalog, and left free the normalization factor of
all the sources within 14$^\circ$ of \psr. All the non-pulsar sources
were modeled with a power law as reported in the 1FGL catalog, while
the two pulsars in the region of interest, PSR J0205+6449 and PSR
J2229+6114, were modeled by a power law with exponential cutoff
according to the data reported in the \f-LAT pulsar catalog
\citep{LATPSRCAT}.

To obtain \f-LAT spectral points we divided our sample into
logarithmically-spaced energy bins (4 bins per decade starting from
100 MeV) and then applied the maximum likelihood method in each
bin. For each energy bin, all point sources, including PSR J0007+7303,
were modeled by a power law with fixed photon index.  From the fit
results we then evaluated the integral flux in each energy bin. If in
an energy bin the source significance is lower than $3\sigma$ we have
evaluated the 95\% integral flux upper limit in that bin.  This method
does not take into account energy dispersion or correlations among the
energy bins.  To obtain the points of the spectral energy
distributions (SEDs) we multiplied the flux in each bin by the
spectrally weighted mean bin energy.\\

\subsubsection{Off-pulse Spectrum}
\label{sec:offpulsespec}
To quantify the off-pulse \g-ray emission we have assumed a point-like
source, modeled with a power law at the pulsar position. We considered
only events in the off-pulse phase interval [0.71 -1.07]. The fitted
power law spectrum is given by:

\begin{equation}
  \frac{dN{(E)}}{dE} = \frac{N (1-\gamma)
    E^{-\gamma}}{E_{max}^{1-\gamma}-E_{min}^{1-\gamma}}
\label{eq_pl_OFF}
\end{equation}
where for this fit $N=1.69\pm 0.40_{stat} \pm 0.18_{sys}
\times10^{-8}$ \pcs, $\gamma = 2.54 \pm 0.14_{stat} \pm 0.05_{sys} $
with $E_{min}=100$ MeV and $E_{max}=100$ GeV.  The estimated integral
flux above 100 MeV is F$_{100} = 1.73 \pm 0.40_{stat} \pm 0.18_{sys}
\times10^{-8}$ \pcssp and the integral energy flux above 100 MeV is
G$_{100} = 7.83 \pm 1.43_{stat} \pm 0.56_{sys} \times10^{-12}$ erg
cm$^{-2}$ s$^{-1}$.  There is no compelling statistical case in favor
of a cutoff in the spectrum, suggesting that the emission is not
magnetospheric in origin. We choose to model the off-pulse emission
with a power law because it has less parameters and gives smaller
statistical errors compared to an exponential law with a cutoff.

Systematics are mainly based on uncertainties in the LAT effective
area derived from the on-orbit estimations, and are of $\leq$ 5$\%$
near 1 GeV, 10$\%$ below 0.1 GeV and 20$\%$ above 10 GeV
\citep{onorbitcal}. We therefore propagate these uncertainties using
modified effective areas bracketing the nominal ones
(P6$\_$V11$\_$DIFFUSE).\\

\subsubsection{On-pulse Spectrum}
\label{sec:onpulse phase averaged}
\begin{figure}[ht]
\begin{center}
\includegraphics[width=0.9\columnwidth]{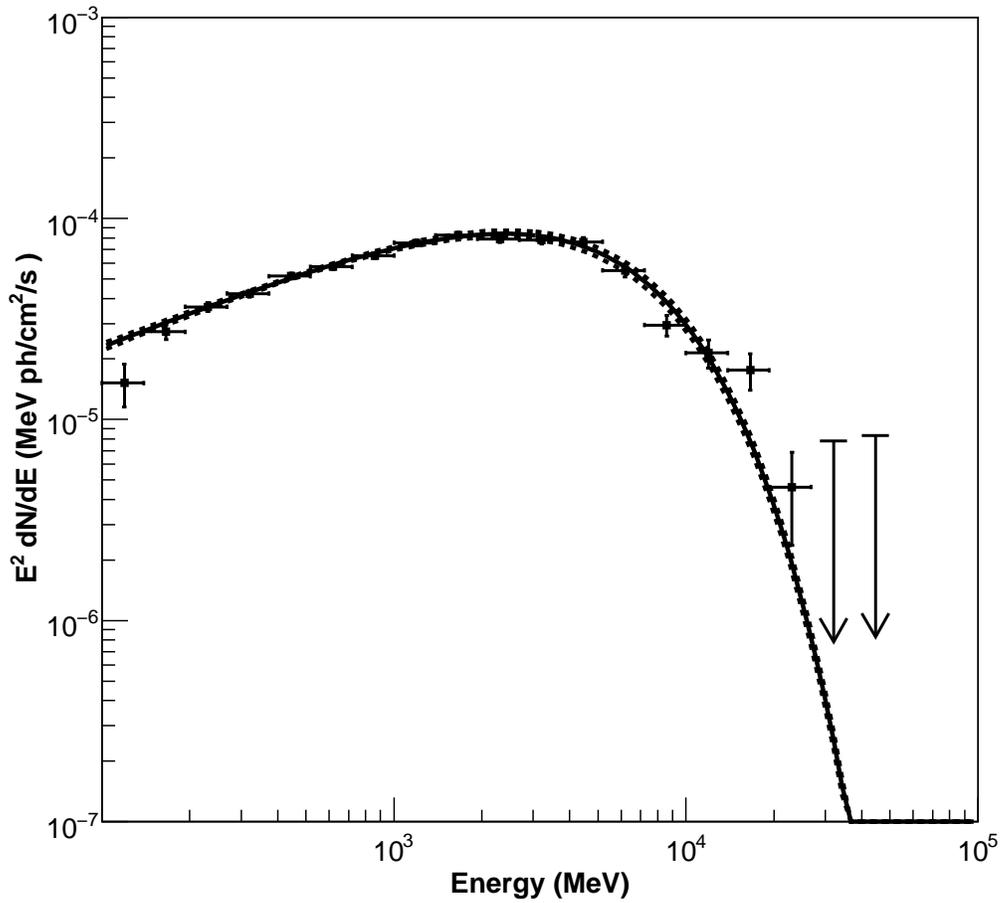}
\caption{On-pulse phase-averaged spectral energy distribution of PSR
  J0007+7303. The solid black line represents the best fit power law
  with exponential cutoff with b=1. Dashed lines represent the 1$\sigma$
  errors in each case.}
\label{fig:cta1_phaseave}
\end{center}
\end{figure}

To account for the off-pulse emission we used the results of the
off-pulse fit, properly rescaled to the on-pulse phase interval, as
a starting point for the pulsed emission analysis. In the model we
considered two sources in the same position, one described as a power
law with the spectral parameters fixed at the values found with the
off-pulse fit and one described by a power law with exponential
cutoff (PLEC) in the form:
\begin{equation}
\frac{dN(E)}{dE} = N_{\circ} \left( \frac{E}{1 \mbox{ GeV}}\right)^{-\Gamma} \exp \left(-\left(
\frac{E}{E_c}\right)^{b}\right) 
\label{eq_plec}
\end{equation}
We set $b=1$ for which the best-fit parameters are $N_{0}$ = ($9.08
\pm 0.20_{stat} \pm 0.54_{sys}) \times$ 10$^{-11}$ cm$^{-2}$ s$^{-1}$
MeV$^{-1}$, $\Gamma$ = ($1.41 \pm 0.23_{stat} \pm 0.03_{sys}$) and
$E_c$ = ($4.04 \pm 0.20_{stat} \pm 0.67_{sys}$) GeV.  The integral
flux above 100 MeV is F$_{100} = $ ($3.95 \pm 0.07_{stat} \pm
0.30_{sys}) \times$ 10$^{-7}$ \pcssp and the corresponding energy flux
is G$_{100} =$ ($4.41 \pm 0.06_{stat} \pm 0.5_{sys}) \times$
10$^{-10}$ erg cm$^{-2}$ s$^{-1}$.  We also fitted the on-pulse
phase-averaged spectrum to different spectral models, power law and
broken power law. Both models can be excluded at the $> 5 \sigma$
level compared to the PLEC model used above.  We have also fitted the
PLEC spectrum while leaving free the exponential index $b$. The $b$
value obtained was lower then 1 but the overall fit did not improve,
thus we adopt the simpler PLEC model for which b=1.  Figure
\ref{fig:cta1_phaseave} shows the results of the on-pulse
phase-averaged spectrum.


\subsubsection{On-pulse Phase-resolved Spectrum}

\begin{figure}[ht]
\begin{center}
 \includegraphics[width=0.8\columnwidth]{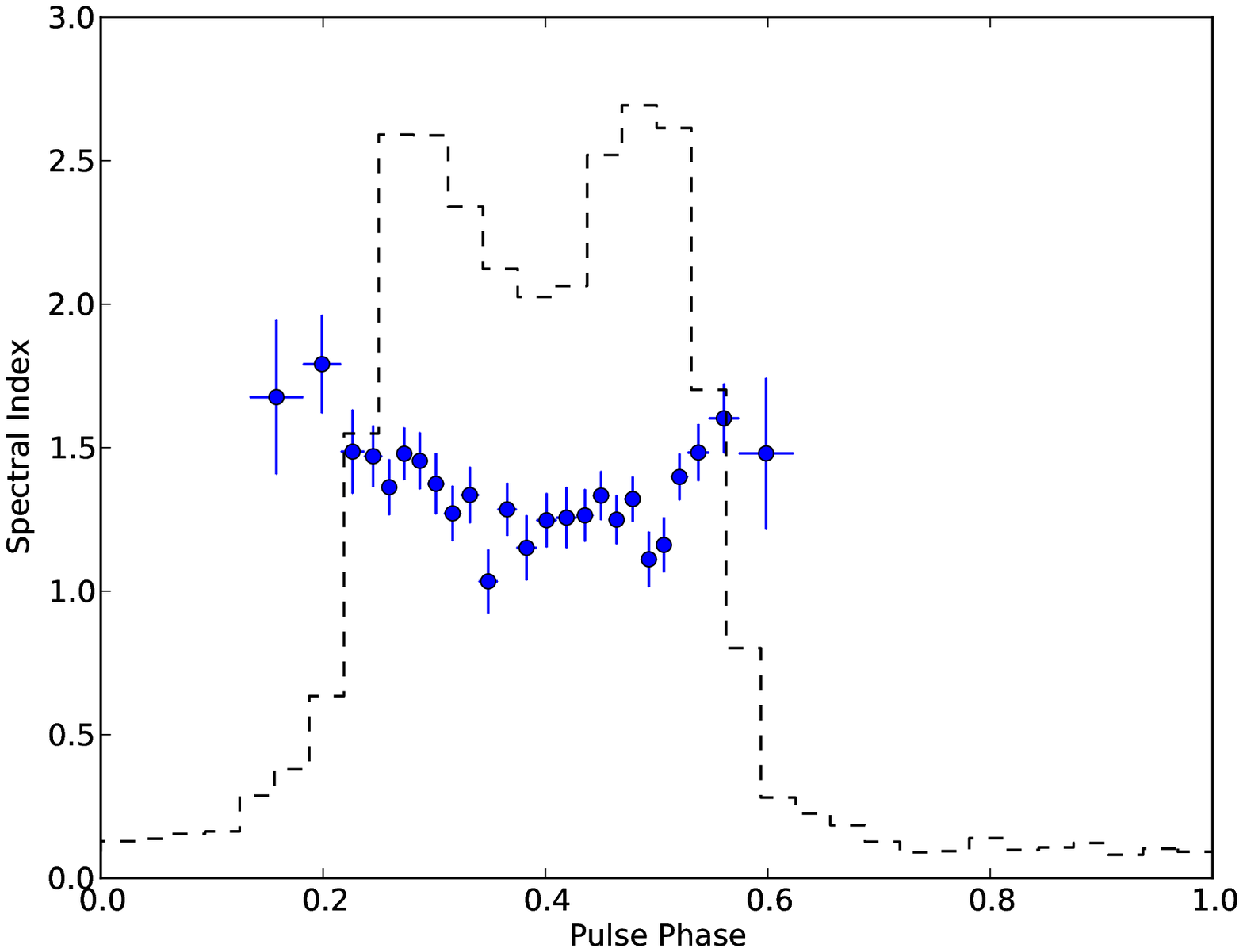}
 \includegraphics[width=0.8\columnwidth]{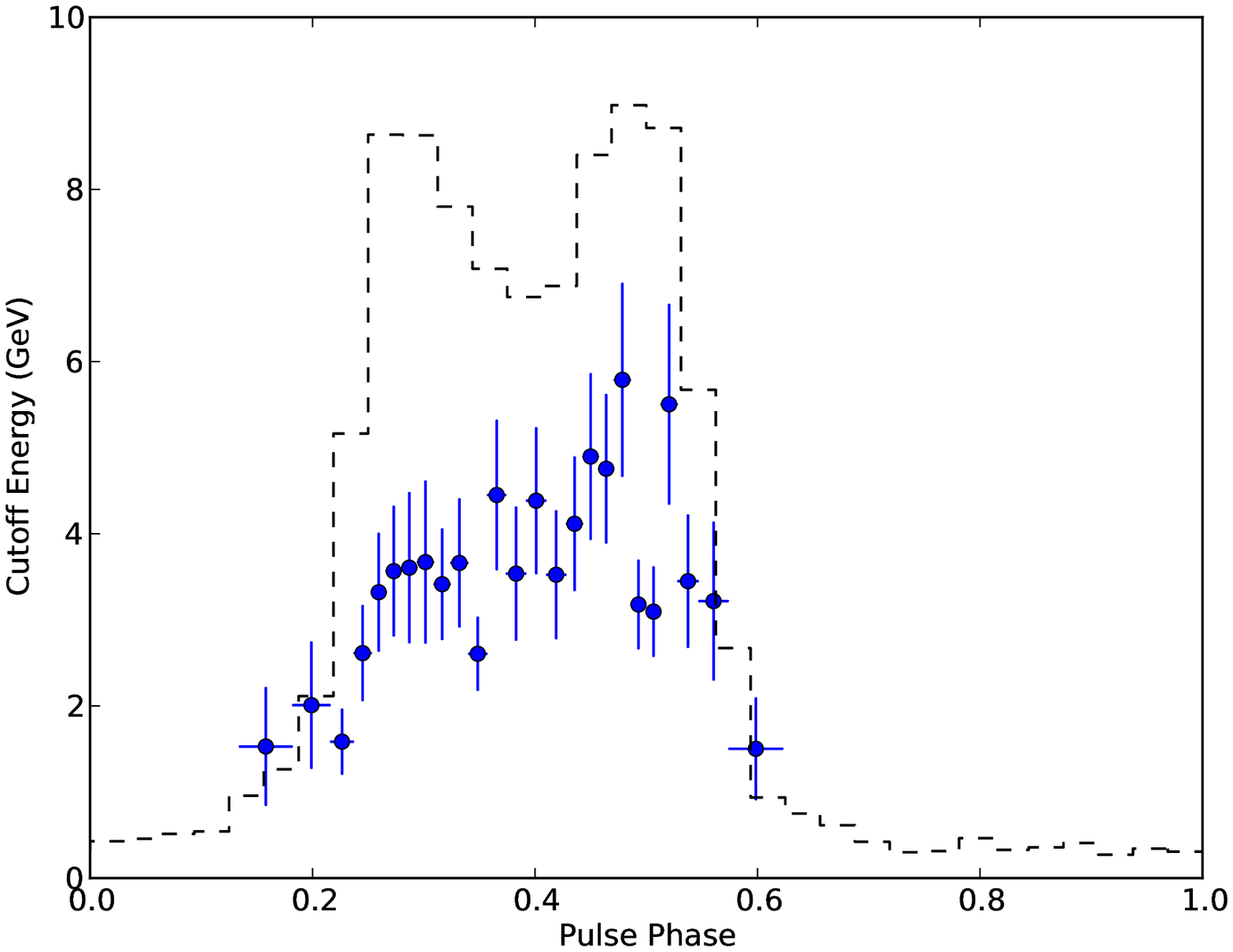}
 \caption{Evolution of photon index (\emph{top}) and energy cutoff
   (\emph{bottom}) above 0.1 GeV as a function of pulse phase from
   fits in fixed-count phase bins of 500 photons per bin. The error
   bars denote statistical errors. For each phase interval (defined in
   Table \ref{tab:phres}) a power law with exponential cutoff has
   been assumed. The dashed histogram represents the LAT light curve
   above 0.1 GeV.}\label{fig:ph_evol}
\end{center}
\end{figure}

To explore the on-pulse phase-resolved spectrum, we divided the pulse
profile in variable-width phase bins, each containing 500 photons
above 100 MeV. These bins were defined by selecting only those events
within an energy-dependent radius of $\theta <$ Max(Min(R$_{max},
\theta_{68}), 0.35\degsp)$ around \psr: the minimum value of
0.35$\degsp$ was selected in order to keep all high-energy photons,
while a maximum radius, $R_{max}=2\deg$, was introduced to reduce the
background contamination at low energies. This choice of binning
provides a reasonable compromise between the number of photons needed
to perform a spectral fit and the length of phase intervals. It should
be short enough to sample fine details on the light curve, while
remaining comfortably larger than the rms of the timing solution.  A
binned maximum likelihood spectral analysis, similar to the analysis
performed in \S \ref{sec:onpulse phase averaged}, was performed in
each phase bin with the exception of fixing the spectral parameters of
all the nearby \g-ray sources and of the two diffuse backgrounds to
the values obtained in the phase averaged analysis, rescaled for the
phase bin width.  Using the likelihood ratio test we can reject the
power law model at a significance level greater than 5$\sigma$ in each
phase interval. Such a model yields a robust fit with a logarithm of
the likelihood ratio greater than 150 in each phase interval. Figure
\ref{fig:ph_evol} shows the evolution of the spectral parameters
across PSR J0007+7303's rotational phase. In particular, the energy
cutoff trend provides a good estimate of the high energy emission
variation as a function of the pulsar phase.  Table \ref{tab:phres}
summarizes the results of the spectral fit in each phase
bin. Variations of both the photon index and the cutoff energy as a
function of rotational phase are apparent. The photon index seems to
show a rough symmetry centered at a phase point half way between the
two peaks with the index increasing (softening) outwards. This is
similar to what has been observed for PSR J1709$-$4429
\citep{3EGRET_psrs}.  The cutoff energy evolves quite differently as a
function of the rotational phase. It increases from a minimum value of
1.5 GeV at $\phi=0.2$ until reaching the first peak where it stays at
a constant value of $\sim 4 $ GeV until the second peak is reached,
after which it seems to fluctuate between 3 and 6 GeV before finally
decreasing to its minimum value of 1.5 GeV at $\phi=0.6$.

\subsection{Light Curve}
\label{sec:LightCurve}
\begin{figure}
\begin{center}
\includegraphics[scale=0.6]{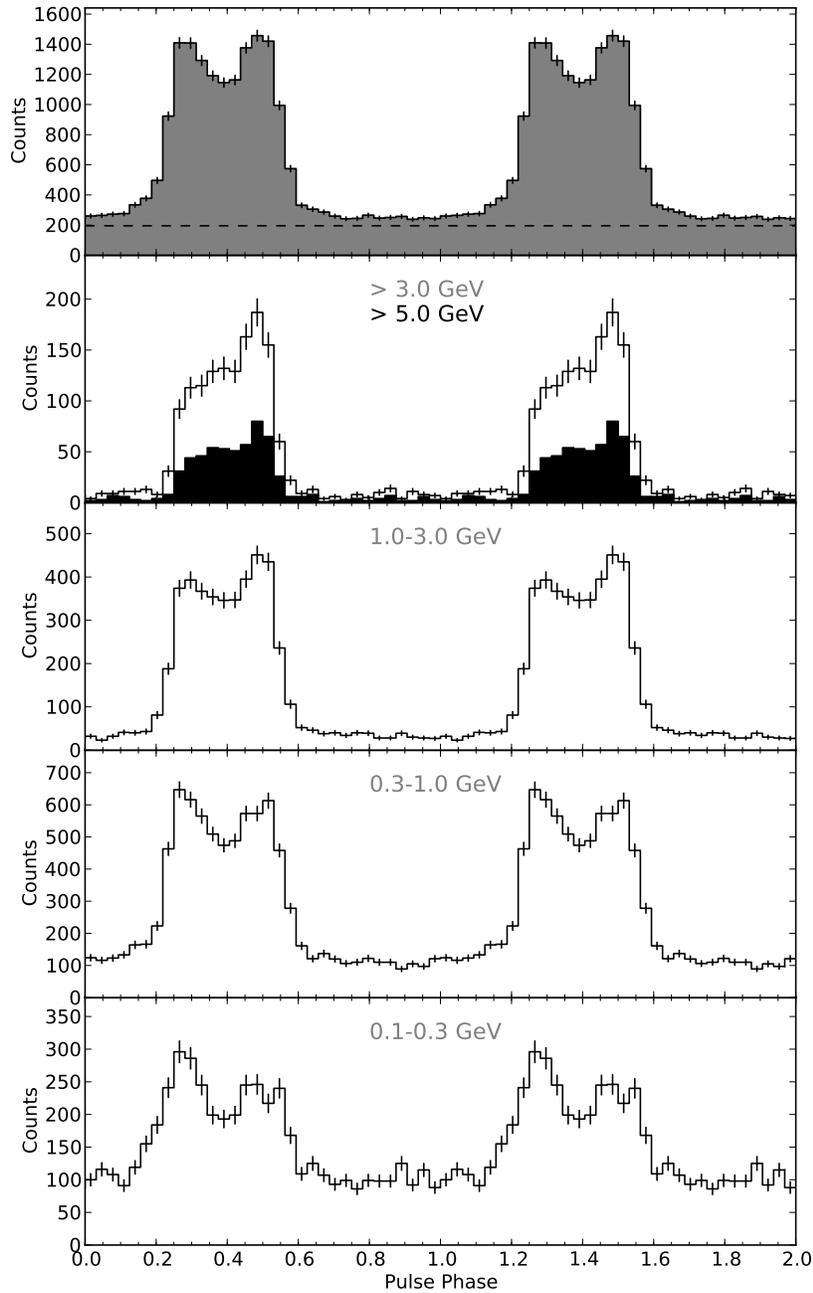}
\caption{The evolution of the pulse profile of PSR J0007+7303 with
  energy. Two rotational periods are shown with a resolution of 32
  phase bins per period. The top panel shows folded light curve for
  energies above 100 MeV. The dashed horizontal line shown in the top
  panel shows the estimated level of the background due to diffuse
  emission (see text for details). The rest of the panels show the
  light curve in exclusive energy bands. The darker histogram on the
  second panel from the top shows the folded light curve for energies
  $>$ 5 GeV} 
\label{fig:lc}
\end{center}
\end{figure}

We investigated the pulsar light curve in different energy bands by
selecting events within $1.6\degr$ of the pulsar position. This
value was selected to maximize the signal-to-background ratio over the
full energy range ($E>100$ MeV). The energy-resolved light curve is
shown in Figure \ref{fig:lc}. The top panel in the Figure shows the
folded light curve for energies above 100 MeV while the rest of the
panels show the light curve in exclusive energy bands. The dashed
horizontal line shown in the top panel shows the estimated level of
the background due to diffuse emission. This background estimate of
$194 \pm 5$ counts/bin was obtained by simulating two years of
data. We used the \lsp Science Tool \texttt{gtobssim} and used for the
input model the best fitted model from \S \ref{sec:onpulse phase
  averaged} but with \psrsp and the off-pulse component (\S
\ref{sec:offpulsespec}) removed from the input
model. 
\newline
The light curve shows two distinct peaks. We
fitted the light curve by a double Gaussian for which the first peak
and second peak are located at $\phi = 0.303 \pm 0.002 $ and $\phi =
0.484 \pm 0.002$ respectively. The separation between the means of the
two peaks is $0.181 \pm 0.003$ in phase. As can be seen from the
Figure there is a significant evolution in the counts ratio of the two
peaks P1/P2.

\subsection{Flux Variability Analysis}
\label{sec:var}
\begin{figure}
\begin{center}
\includegraphics[scale=0.8,angle=0]{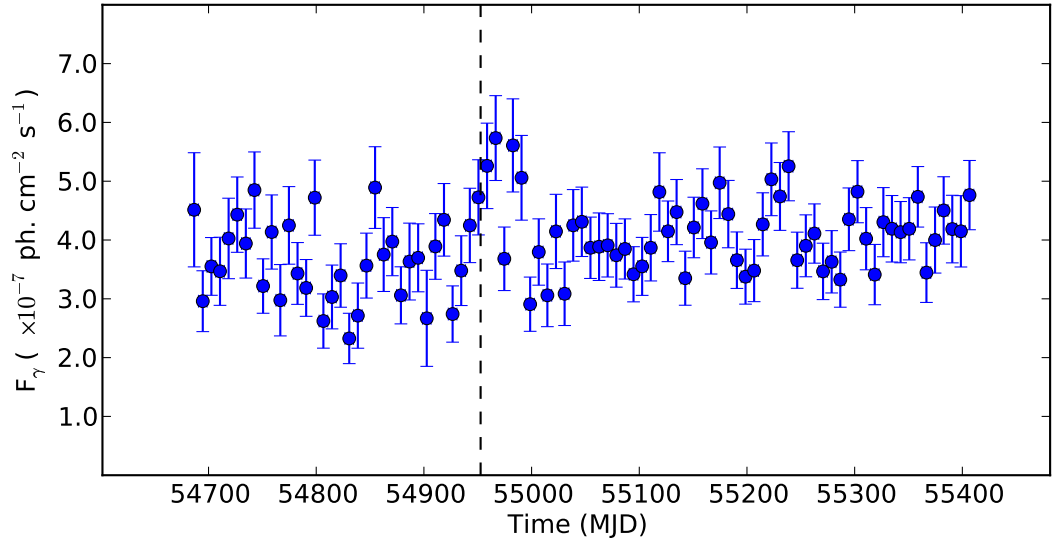}
\caption{Flux ($> 100 $ MeV) of PSR J0007+7303 as a function of time
  in 8-day time bins. The flux shows no evidence for variability. The
  dashed vertical line marks the time of the glitch.  }
\label{fig:variability}
\end{center}
\end{figure}

To check for long-term stability in the flux we performed likelihood
analysis similar to that in \S \ref{sec:spectral} but in 8-day time
bins. Figure \ref{fig:variability} shows the resulting fluxes. The
length of the time bin was selected to allow for the accumulation of
enough statistics to guarantee a good likelihood fit.  To look for
flux variability from the source we adopt the method outlined in
\citep{Fermicatalog}. The source shows no variability on this time
scale. Similar analyses were performed for 16, 32, and 64-day time
bins, no significant modulation was found.

To check for any change in the spectrum of the pulsar due to the
glitch we split the data in two bins around the glitch. The pre-glitch
epoch spans the time range 54682.68 -- 54952.652 (MJD), while the
post-glitch data spans the time range 54952.652 -- 55412.65 (MJD).  We
performed a likelihood analysis similar to that in \S \ref{sec:var}
in these two time bins. In Table \ref{table:glitch} we show the
spectral fits for the pulsar in these two epochs. The flux and
spectral parameters are in good agreement for the two epochs. No
change in integral flux above 100 MeV is seen.

Recent variability detected in the Crab appears to come from the
Nebula, not the pulsar \citep{CrabFlare}. Since \psrsp is among the
younger pulsars and has a complex PWN/SNR associated with it, it is
reasonable to ask whether analogous variability occurs in this source.
While \S \ref{sec:offpulse} has shown strong evidence for an
off-pulse component, tentatively extended, there are not enough
statistics to explore variability in this component at all, and
certainly not on the time scales detected in the Crab.

\section{Discussion}

\subsection{Geometrical Constraints from Light Curve Modeling}
\label{sec:LC_modeling}
The pulsar emission mechanism is not well understood, and the
distributions of pulsar magnetic inclination angles, efficiencies, and
other physical characteristics are largely unknown. There are several
competing emission models, of which the resulting pulse profiles
depend on the geometry of the system, defined by the emission zone
location and size, observer viewing angle, and magnetic inclination
angle. Fitting the profiles of these models to observed light curves
can provide insight on the true emission and viewing geometries, for
example leading to better constraints on luminosity and efficiency
through calculation of the flux correction factor $f_{\Omega}$
(equation 4 of \cite{Watters09}). The geometry of a given system
likely also contributes to the detectable presence or absence of radio
and perhaps X-ray emission. For example, if the inclination and
observer angles are very different, we could see a $\gamma$-ray-only
pulsar, as the narrower radio beam may not cross our line of sight. To
determine the geometry of the CTA 1 pulsar and distinguish between
emission models, we compared the LAT light curve of PSR J0007+7303
with the predicted light curves from geometrical representations of
two standard high-energy pulsar emission models, the outer gap (OG)
\citep{Romani1995} and slot gap (SG, also referred to as the two-pole
caustic or TPC) \citep{Muslimov04} models. These models were
considered within the context of the vacuum retarded dipole magnetic
field.

The light curves were simulated as in \cite{Dyks2004}, with the
geometry modified to represent each emission region. The OG lies along
the last open field lines between the null charge surface and the
light cylinder, $R_{\mathrm{lc}} = c/\Omega$, with $r_{\mathrm{cyl}}
\, \leq \, 1.0R_{\mathrm{lc}}$ as large as possible for each
geometrical configuration. The SG extends from the surface to
$r_{\mathrm{cyl}} = 0.95R_{\mathrm{lc}}$. In both geometries, the
maximum emission altitude $r_{\mathrm{max}}$ was treated as a model
parameter. We note that our representation of the SG differs from that
of the TPC model in \cite{Romani2010}, as we allow emission out to
much higher altitudes (their emission is cut off at $r_{\mathrm{cyl}}
= 0.75R_{\mathrm{lc}}$). The TPC geometry is therefore
treated as a subset of the SG, rather than being considered as a
completely separate model. The SG model has emission throughout the
gap, while the OG emission occurs only along the field line at the
gap's innermost edge. In the simulations, the vacuum retarded dipole
field is assumed in the observer's frame and then transformed to the
co-rotating frame (CF). Photons are emitted tangent to the field in
the CF, prior to the aberration calculation. A constant emissivity is
assumed along the field lines in the CF. The special relativistic
aberration leads to a bunching in pulse phase, or caustics, on the
trailing side of the pulse, producing peaks in the light curvyes. For a
given inclination angle $\alpha$, gap width $w$ in open volume units
($r_{\mathrm{ovc}}$, as in \cite{Dyks2004}), and maximum emission
radius $r_{\mathrm{max}}$ in units of $R_{\mathrm{lc}}$, the code
produces light curves at all observer angles $\zeta$.

The LAT light curve of PSR J0007+7303 has 32 bins and
a background count level of 195 or 246 counts/bin, depending on
whether the off-peak emission discussed in \S2.2 is (respectively)
magnetospheric or due to a PWN. We compared this light curve with
light curves simulated from a representative sample of all possible
geometries and rebinned to match the observed light curve. Our models
had five free parameters, $\alpha$, $\zeta$, $w$, $r_{\mathrm{max}}$,
and a phase shift $\Delta \phi$ introduced in order to best match the
LAT light curve.
We considered values of $\alpha$ between $0^{\circ}$ and $90^{\circ}$
and $\zeta$ between $0.5^{\circ}$ and $89.5^{\circ}$, each with
resolution of $1^{\circ}$; values of width $0 \leq w \leq 0.3$ with
resolution 0.01$\,r_{\mathrm{ovc}}$; and maximum emission radii $0.7
\leq r_{\mathrm{max}} \leq 2$ for the SG and $0.9 \leq r \leq 2$ for
the OG, with resolution 0.1$\,R_{\mathrm{lc}}$ in both cases. For each
model, the emission altitude is limited by Min($r_{\mathrm{max}},
r_{\mathrm{cyl}}$)---$r_{\mathrm{max}}$ therefore may be larger than
$r_{\mathrm{cyl}}$, but emission ceases at the cylindrical radius if
not at the maximum emission radius. The phase shift $0^{\circ} \leq
\Delta \phi \leq 360^{\circ}$ is arbitrary because PSR J0007+7303 is
radio-quiet within the flux limits achieved thus far by radio
telescopes; were the pulsar radio-loud, the shift would be constrained
to be at most equal to the phase lag between the radio and
$\gamma$-ray peaks \citep{Dyks2004}.


To find the best-fit parameters of each model, we used a Markov Chain
Monte Carlo (MCMC) maximum likelihood routine as described in
\cite{Verde2003} that explored the parameter space. We calculated the
$\chi^2$ between the LAT light curve and model light curve and used
Wilks' Theorem, $\Delta\,$ln$(L) = -\Delta \chi^2/2$, to guide the
Markov chains toward regions of high likelihood. We also used the
$\Delta \chi^2$ test to find $3\sigma$ confidence intervals on each
parameter. For this pulsar, the best-fit light curve has a very
large $\chi^2$. This is expected, as a) the pulsar is bright and its
light curve has relatively small error bars and b) the light curve
simulations result from geometrical models of possible but simplified
pulsar magnetospheres, rather than from a well-understood physical
model. When finding confidence intervals in steps of $\Delta \chi^2$
from a large initial $\chi^2$ value, the intervals will appear
artificially small. We therefore re-scale the $\chi^2$ values found
for all sets of parameters in the MCMC chains so that the minimum
reduced $\chi^2 = N_{\mathrm{dof}} = 27$ (the scale factor is
therefore $N_{\mathrm{dof}} / \chi^2_{\mathrm{min}}$). The confidence
intervals are then found with these modified $\chi^2$ values. In this
way, we find parameter ranges that may give rise to the high-energy
light curve of PSR J0007+7303.

\begin{figure}[ht]
\begin{center}
\includegraphics[scale=0.8]{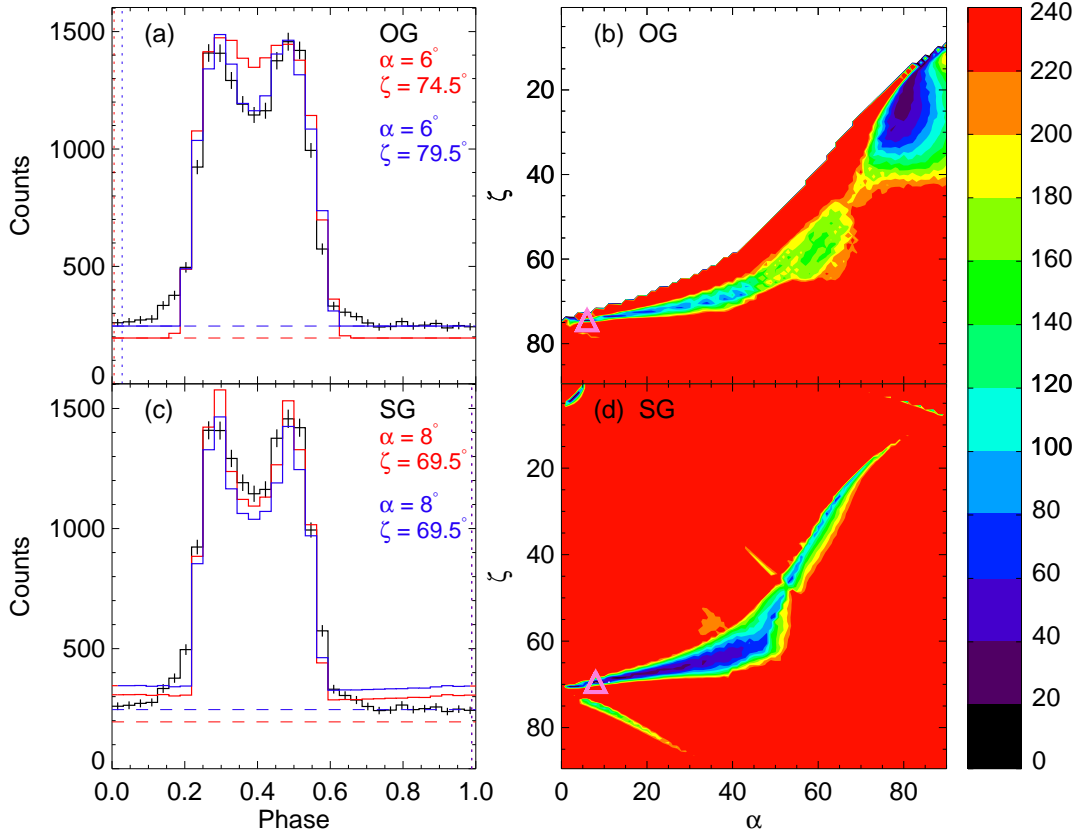}
\caption{The best fitting model light curves for each model, red
  indicating 195 counts/bin and green 246 counts/bin as the background
  level, and the reduced $\chi^2_{\mathrm{r}} =
  \chi^2/N_{\mathrm{dof}}$ contours in $\alpha$--$\zeta$ parameter
  space for the case of a background of 195 counts/bin. ({\textit a})
  Best fitting outer gap light curve (red/blue) superposed on the LAT light
  curve of the \psrsp pulsar (black); the best fit parameters are
  given in the text. The horizontal dashed lines represent the
  estimated background levels, while the vertical
  dotted lines mark the location of the magnetic pole in the context
  of the outer gap model for the listed sets of parameters. ({\textit b})
  $\chi^2_r$ contours for outer gap light curves in which $w$ and $r$
  have been fixed at the best-fit parameters. The filled contours show
  $\chi^2$ in increments of 20 for all $\alpha$ and $\zeta$. The best
  fit is marked by the pink triangle.
The $3\, \sigma$ confidence intervals for each parameter are given in
  Table~\ref{table:model_fits}. ({\textit c}) Same as ({\textit a}),
  for the slot gap model. ({\textit d}) Same as ({\textit b}), for the
  slot gap model.}
  \label{fig:model_fits}
\end{center}
\end{figure}

For the case where the off-peak emission is assumed to
be magnetospheric (corresponding to a background level of 195
counts/bin), the best-fit parameters are ($\alpha$, $\zeta$, $w$,
$r$, $\Delta \phi$) = (6$^{\circ}$, 74.5$^{\circ}$, 0.04$r_{\,
\mathrm{ovc}}$, 1.0$\,R_{\mathrm{lc}}$, 2$^{\circ}$) with $\chi^2/27 =
22.5$ for the OG model and (8$^{\circ}$, 69.5$^{\circ}$, 0$\,
r_{\mathrm{ovc}}$, 1.9$\,R_{\mathrm{lc}}$, -4$^{\circ}$) with
$\chi^2/27 = 11.3$ for the SG model. Almost identical
parameters are found for the case where the off-peak emission is
assumed to originate from a PWN. The confidence intervals,
$f_{\Omega}$ values, and reduced $\chi^2$ for the best fit in each
interval are given in Table~\ref{table:model_fits}. The absolute best
fit is found with the OG model using the higher background level.
Figure~\ref{fig:model_fits} shows the LAT light curve superposed with
the best-fit model light curves, as well as reduced $\chi^2$
contours in $\alpha$ and $\zeta$ with $w$ and $r$ fixed at the
best-fit values.
To be complete, we also fit the light curve with the
TPC model, the results of which are not shown here; we find this
lower-altitude subset of the SG model cannot reproduce the sharp
double peaks with the correct spacing anywhere in parameter space.
We therefore find that both outer magnetosphere models produce light
curves consistent with that which is observed.  For magnetospheric
off-peak emission, the SG is preferred over the OG; its ability to
reproduce the light curve is seen clearly by comparing the red curves
in panels ({\textit a}) and ({\textit c}) of
Figure~\ref{fig:model_fits}. The reverse is true under the assumption
of off-peak emission from a wind nebula, shown by the blue curves.
Under the assumption of magnetospheric off-pulse emission, the SG
misses some emission in the wings and has too high a background level;
this effect is magnified when we assume instead a PWN origin of this
emission. Such details may be modified with a more physical emission
model, for example by including azimuthal asymmetry in the
accelerating electric field from offset polar caps, which leads to a
decrease in the off-peak emission \citep{Harding2011}. The OG
characteristically does not reproduce the wings or higher emission in
the off-peak phases when the background is derived assuming pulsar
emission in the off-peak; by definition it matches the background well
when the background is instead found assuming the emission is not from
the pulsar itself.

Regardless of the background counts used, the best-fit
values of $\alpha$, and $\zeta$ for the OG and SG models are far
apart, with $|\zeta-\alpha| >$60 degrees. The radio beam would need a
width of that order in order for the radio pulse to be detectable. For
the parameters of \psrsp a model estimate of beam width is $<$ 10
degrees \citep{Story07}. Thus the preference for OG and SG model fits
over that of TPC, along with the fitted parameter values,
self-consistently offers a satisfactory explanation for non-detection
of any radio pulse. A deep search using GBT \citep{Halpern2004}
yielded an upper limit that remains the lowest limiting flux for any
pulsar position. The upper limit on the pseudo-luminosity of $L_{1400}
\simeq 0.02 $ mJy kpc$^2$ is lower than the lowest pseudo-luminosity
measured for a radio pulsar ($L_{1400}\,\simeq \,0.035$ mJy kpc$^2$
for PSR J1907+0602 \citep{1907}) and is thus restrictive enough to
support the radio-quiet designation. Conversely, if one were to adopt
the TPC model for this source then the nearly-equal values of $\alpha$
and $\zeta$ would suggest that a radio pulse should have been found.

X-ray pulsations were recently detected from the CTA 1 pulsar using
the same LAT ephemeris used in the timing and light curve analysis of
this paper \citep{Lin2010,Caraveo2010}.  We did not include any X-ray
information in our fits, but can consider whether or not our results
make sense in the multiwavelength picture.
The peak in the X-ray light curve shows a significant thermal
component, suggestive of a hot spot on the surface. The authors
estimate the size of the hot spot to be $\sim 100\,$m in radius,
larger than the polar cap in the case of a dipole model for the
pulsar; it is not clear from the data where the hot spot is located.
The X-ray peak occurs at $\phi \sim 0.25$--0.3 ($\sim
90^{\circ}$--110$^{\circ}$) prior to the first peak in the
$\gamma$-ray light curve. Our HE models predict a similar location of
the magnetic pole: the first $\gamma$-ray peak is at $90^{\circ}$, and
in all our model fits the pole lies $36^{\circ}$--$94^{\circ}$ before
this peak; taking only the SG and best OG fits places the pole
$94^{\circ}$--$88^{\circ}$.
Our best-fit models are therefore consistent with the thermal X-ray
emission originating near the magnetic pole.

\subsection{Thermal Component in X-rays and Neutron Star Cooling}

\psrsp is among the neutron stars (NS) where the theory of cooling
confronts observation. It held that distinction before launch of \f,
based on the candidate NS identified at that time \citep{Halpern2004},
and it continues to be treated as an exceptional case in the cooling
literature \citep{Page2009}.  Those references cite additional
literature on heat loss models which describe how observable surface
temperature or thermal flux declines with age.  X-ray or extreme ultra
violet surface emission in young, hot NS provides the observational
test.  The NS age and surface temperature must be known or
constrained, that is, a limit may be useful if corrections can be
characterized at least as to sign if not magnitude.  Only
comparatively young NS within a few kpc provide useful
constraints. The only pulsar both younger and closer than \psrsp is
the Vela pulsar.  Before \f, the age of \psrsp was equated to the
estimated CTA1 SNR age and thermal flux was estimated from the X-ray
emission of the identified point-like NS candidate \citep{Halpern2004}.
From this it was understood that the putative pulsar in CTA1 was cool
for its age, in comparison with both theory and other young NS.  This
led to theoretical effort to understand why this particular NS had
cooled rapidly.

\fermis results impact context information that enters into the
analysis of the cooling question.  The pulsar ephemeris affects the
X-ray analysis used to extract the thermal flux.  The \fermis results
have not greatly altered best-estimates of key quantities but provide
a new network of interlocking constraints that give greater robustness
of the observational context information, as follows.  (1) The spin
period $P$ and its derivative $\dot{P}$ give spindown age of
$\sim$13,900 yr for the pulsar.  This age estimate is well within the
broader range of historical estimates for the SNR age, 5,000-15,000
yr, and nearly identical to one pre-launch estimate of $1.3 \times
10^4 \, d_{1.4} $ yr \citep{Slane2004}.  The formal propagated error
in the spindown age estimate is negligible.  The principal issue is
validity of the dipole approximation formula but it clearly reinforces
the age estimates derived other ways. (2) The same observations ($P$,
$\dot{P}$) also yield spindown energy loss independently of distance.
There are no inconsistencies between this energy budget and that of
the pulsar plus SNR, using the accepted distance, D, of $1.4 \pm 0.3$
kpc, hence the pre-\fermis distance estimate used is not in conflict
with new information and the distance scale factor in the age estimate
just cited is consistent with unity. (3) The pulsar ephemeris has now
been used to detect X-ray pulsations using XMM \citep{Lin2010,
  Caraveo2010}.  The X-ray spectrum can be divided into pulsed and
unpulsed components. While total X-ray flux is consistent with levels
measured in pre-\f\ work, the part now assigned to a DC thermal
component emitted from the surface is reduced by at least a factor
that equates to the unpulsed fraction.  Since D is unchanged, the
revised bound on NS surface temperature remains as in earlier
estimates or perhaps even slightly reduced.  One may consider
modifications due to the interstellar absorbing column and to the
neutron star atmosphere causing the surface flux to depart from
blackbody.  These considerations properly enter into many other NS
used for comparison with cooling and it is beyond scope of this paper
to reanalyze the entire X-ray discussion. It is left to future
research but provisionally the pre-launch thermal luminosity stands.
(4) In this same connection a further point to note is that the
off--pulse \g-ray flux, discovered and quantified in \S
\ref{sec:offpulse} of the present paper, cannot be thermal flux from
the neutron star and must be taken as magnetospheric or PWN emission.
In either case, that same magnetosphere or PWN component could extend
downward in energy to X-rays, and in principle could provide further
downward adjustment to the thermal X-ray flux from the star.  This is
left as an adjustment of unknown magnitude but known sign; it can only
require the star to cool faster than the estimate obtained by
neglecting it. (5) From $P$ and $\dot{P}$ one also obtains an estimate
of the stellar dipole field, B $\sim10^{13}$G. \psrsp therefore falls
among the most highly magnetic neutron stars, magnetars excepted.
Cooling scenarios whereby a strong magnetic field could modify the
cooling curve have been described in earlier literature (for a summary
see for example \cite{Yakovlev2001}).  The high value of B now
established for this pulsar means mechanisms whereby high B enhances
cooling may merit further attention.  (6) The previous section shows
how model fits for OG and SG models favor small $\alpha$ and large
$\zeta$, and could explain the absence of radio pulsations.

The first four items in the list mean that if \psrsp was an outlier
relative to models before \textit{Fermi} and relative to
pre-\textit{Fermi} theoretical understanding, it has become slightly
more egregious relative to that prior theoretical understanding.
However theory has also advanced, largely from observation of cooling
in the central star in Cas A \citep{Page2009}.  Data from that source
are now the most definitive and constraining of any un-recycled NS.
Fitting Cas A has given prominence to NS cooling models involving
Cooper pairing contributions.  Points (5) and (6) in the list
regarding magnetic field strength and geometry may provide guidance to
theory.  Sufficiently strong magnetic fields can affect cooling
\citep{Yakovlev2001} and might affect heat flow in the star.  Also the
particular geometry in \psrsp with low $\alpha$ and high $\zeta$ that
operates against radio pulse detection could also lead to a
misleadingly low thermal X-ray flux if the strong field conveyed
internal heat flow preferentially to the magnetic polar regions.
Then, a Lambertian emission pattern from the hotter poles would be
anisotropically beamed away from an observer at high $\zeta$.  Further
X-ray observations, combined with multi-wavelength analysis applied
and field geometry modeling may shed further light on the thermal
luminosity.  This could be undertaken comparatively with other young,
nearby pulsars such as the “Dragonfly” pulsar, PSR J2021+3651, which
has estimated age 17 kyr, distance 2.1 kpc, dipole field $3 \times
10^{12}$ G, and has radio pulses.  Such comparisons might bring out
the role of the magnetic field strength and geometry in NS cooling.

\section{Summary} 
\psrsp is among the brightest \g-ray pulsars \citep{LATPSRCAT}.  It
also is part of an interesting PWN and SNR complex that is still
young. We have exploited greatly-improved cumulative statistics from
two years of \f-\lsp data to investigate questions that can be pursued
only on brighter \g-ray pulsars. Interesting aspects of the source
include its having had a major glitch during the \f\ observing period,
its being among the most strongly magnetic of neutron stars that are
not magnetars, and its status as a prime testbed for neutron star
cooling theory, a topic now receiving renewed attention because of the
cooling observed in X-rays in Cas A.

Pulsar phase dependence of the light curve and spectrum have been
investigated and the source has been compared both with other
well-studied pulsars (notably PSR J1709-4429) and also with standard
magnetospheric geometry models, revealing a clear preference for
models where emission occurs high in the magnetosphere; particularly
the SG model.  Glitch parameters have been extracted.  We have
conducted a systematic search for long-term variability in the system,
with negative results.  Neither a change associated with the glitch
nor flaring such as has recently been seen in the younger Crab Nebula
\citep{CrabFlare} has been found.  However, off-pulse emission has
finally been detected at high confidence and there is evidence that it
is extended; hence there is now a new \g-ray component in the overall
source, potentially a PWN although the possibility that it originates
inside the magnetosphere is not strongly excluded.  The variability of
that off-pulse source is a subject to be pursued as \f\ continues to
accumulate data.  We have described how the parameters emerging from
the \g-ray analysis (ephemeris, spindown energy loss, age, distance,
and magnetic field) and from follow-on X-ray studies affect
understanding of the cooling history, reinforcing the conclusion that
the surface of this NS is cool for its age.  This now needs to be
followed up with additional X-ray analysis to further constrain the
surface X-ray luminosity.

\begin{thebibliography}{35}
\expandafter\ifx\csname natexlab\endcsname\relax\def\natexlab#1{#1}\fi

\bibitem[{{Abdo} {et~al.}(2010{\natexlab{a}}){Abdo}, {Ackermann}, {Ajello},
  {Allafort}, {Antolini}, {Atwood}, {Axelsson}, {Baldini}, {Ballet},
  {Barbiellini}, \& et~al.}]{Fermicatalog}
{Abdo}, A.~A. {et~al.} 2010{\natexlab{a}}, \apjs, 188, 405

\bibitem[{{Abdo} {et~al.}(2011){Abdo}, {Ackermann}, {Ajello}, {Allafort},
  {Baldini}, {Ballet}, {Barbiellini}, {Bastieri}, {Bechtol}, {Bellazzini},
  {Berenji}, \& et~al.}]{CrabFlare}
---. 2011, Science, 331, 739

\bibitem[{{Abdo} {et~al.}(2009{\natexlab{a}}){Abdo}, {Ackermann}, {Ajello},
  {Ampe}, {Anderson}, {Atwood}, {Axelsson}, {Bagagli}, {Baldini}, {Ballet}, \&
  et~al.}]{onorbitcal}
---. 2009{\natexlab{a}}, Astroparticle Physics, 32, 193

\bibitem[{{Abdo} {et~al.}(2010{\natexlab{b}}){Abdo}, {Ackermann}, {Ajello},
  {Atwood}, {Axelsson}, {Baldini}, {Ballet}, {Barbiellini}, {Baring},
  {Bastieri}, \& et~al.}]{psrcat}
---. 2010{\natexlab{b}}, \apjs, 187, 460

\bibitem[{{Abdo} {et~al.}(2010{\natexlab{c}}){Abdo}, {Ackermann}, {Ajello},
  {Atwood}, {Axelsson}, {Baldini}, {Ballet}, {Barbiellini}, {Baring},
  {Bastieri}, \& et~al.}]{LATPSRCAT}
---. 2010{\natexlab{c}}, \apjs, 187, 460

\bibitem[{{Abdo} {et~al.}(2010{\natexlab{d}}){Abdo}, {Ackermann}, {Ajello},
  {Baldini}, \& {others}}]{1907}
{Abdo}, A.~A., {Ackermann}, M., {Ajello}, M., {Baldini}, L., \& {others}.
  2010{\natexlab{d}}, \apj, 711, 64

\bibitem[{{Abdo} {et~al.}(2008){Abdo}, {Ackermann}, {Atwood}, {Baldini},
  {Ballet}, {Barbiellini}, {Baring}, {Bastieri}, \& {others}}]{CTA1}
{Abdo}, A.~A. {et~al.} 2008, Science, 322, 1218, (CTA1)

\bibitem[{{Abdo} {et~al.}(2009{\natexlab{b}}){Abdo}, {Ackermann}, {Atwood},
  {Baldini}, {Ballet}, {Barbiellini}, {Baring}, {Bastieri}, \& {others}}]{BSP}
---. 2009{\natexlab{b}}, Science, 325, 840, (Blind Search Pulsars)

\bibitem[{{Abdo} {et~al.}(2010{\natexlab{e}}){Abdo}, {Ajello}, {Antolini},
  {Baldini}, {Ballet}, {Barbiellini}, {Baring}, {Bastieri}, {Bechtol},
  {Bellazzini}, {Berenji}, {Bonamente}, {Borgland}, {Bouvier}, {Bregeon},
  {Brez}, {Brigida}, {Bruel}, {Buehler}, {Burnett}, {Buson}, {Caliandro},
  {Camilo}, {Caraveo}, {{\c C}elik}, {Chekhtman}, {Cheung}, {Chiang},
  {Ciprini}, {Claus}, {Cognard}, {Cohen-Tanugi}, {Dermer}, {de Palma}, {Digel},
  {Silva}, {Drell}, {Dubois}, {Dumora}, {Favuzzi}, {Ferrara}, {Fortin},
  {Frailis}, {Freire}, {Fukazawa}, {Funk}, {Fusco}, {Gargano}, {Gehrels},
  {Germani}, {Giglietto}, {Giordano}, {Giroletti}, {Glanzman}, {Godfrey},
  {Grenier}, {Grondin}, {Grove}, {Guillemot}, {Guiriec}, {Hadasch}, {Hanabata},
  {Harding}, {Hays}, {J{\'o}hannesson}, {Johnson}, {Johnson}, {Johnson},
  {Johnston}, {Kamae}, {Katagiri}, {Kataoka}, {Keith}, {Kerr},
  {Kn{\"o}dlseder}, {Kramer}, {Kuss}, {Lande}, {Latronico}, {Lee},
  {Lemoine-Goumard}, {Longo}, {Loparco}, {Lott}, {Lubrano}, {Makeev},
  {Manchester}, {Marelli}, {Mazziotta}, {Mitthumsiri}, {Mizuno}, {Moiseev},
  {Monte}, {Monzani}, {Morselli}, {Moskalenko}, {Murgia}, {Nakamori}, {Nolan},
  {Norris}, {Noutsos}, {Nuss}, {Ohsugi}, {Okumura}, {Orlando}, {Ormes},
  {Ozaki}, {Panetta}, {Parent}, {Pelassa}, {Pepe}, {Pesce-Rollins}, {Piron},
  {Rain{\`o}}, {Razzano}, {Reimer}, {Reimer}, {Reposeur}, {Ripken}, {Romani},
  {Sadrozinski}, {Sander}, {Saz Parkinson}, {Sgr{\`o}}, {Siskind}, {Smith},
  {Smith}, {Spandre}, {Spinelli}, {Strickman}, {Suson}, {Takahashi}, {Tanaka},
  {Thayer}, {Thayer}, {Theureau}, {Thompson}, {Thorsett}, {Tibaldo}, {Tibolla},
  {Torres}, {Tosti}, {Tramacere}, {Usher}, {Vandenbroucke}, {Vasileiou},
  {Vitale}, {Waite}, {Wang}, {Weltevrede}, {Winer}, {Yang}, {Ylinen}, \&
  {Ziegler}}]{3EGRET_psrs}
---. 2010{\natexlab{e}}, \apj, 720, 26

\bibitem[{{Ackermann} {et~al.}(2011){Ackermann}, {Ajello}, {Baldini}, {Ballet},
  {Barbiellini}, {Bastieri}, {Bechtol}, {Bellazzini}, {Berenji}, {Bloom},
  {Bonamente}, {Borgland}, {Bouvier}, {Bregeon}, {Brez}, {Brigida}, {Bruel},
  {Buehler}, {Buson}, {Caliandro}, {Cameron}, {Camilo}, {Caraveo},
  {Casandjian}, {Cecchi}, {{\c C}elik}, {Charles}, {Chekhtman}, {Cheung},
  {Chiang}, {Ciprini}, {Claus}, {Cognard}, {Cohen-Tanugi}, {Conrad}, {Dermer},
  {de Angelis}, {de Luca}, {de Palma}, {Digel}, {Silva}, {Drell}, {Dubois},
  {Dumora}, {Favuzzi}, {Focke}, {Frailis}, {Fukazawa}, {Funk}, {Fusco},
  {Gargano}, {Germani}, {Giglietto}, {Giommi}, {Giordano}, {Giroletti},
  {Glanzman}, {Godfrey}, {Grenier}, {Grondin}, {Grove}, {Guillemot}, {Guiriec},
  {Hadasch}, {Hanabata}, {Harding}, {Hayashi}, {Hays}, {Hobbs}, {Hughes},
  {J{\'o}hannesson}, {Johnson}, {Johnson}, {Johnston}, {Kamae}, {Katagiri},
  {Kataoka}, {Keith}, {Kerr}, {Kn{\"o}dlseder}, {Kramer}, {Kuss}, {Lande},
  {Latronico}, {Lee}, {Lemoine-Goumard}, {Longo}, {Loparco}, {Lovellette},
  {Lubrano}, {Lyne}, {Makeev}, {Marelli}, {Mazziotta}, {McEnery}, {Mehault},
  {Michelson}, {Mizuno}, {Moiseev}, {Monte}, {Monzani}, {Morselli},
  {Moskalenko}, {Murgia}, {Nakamori}, {Naumann-Godo}, {Nolan}, {Noutsos},
  {Nuss}, {Ohsugi}, {Okumura}, {Ormes}, {Paneque}, {Panetta}, {Parent},
  {Pelassa}, {Pepe}, {Pesce-Rollins}, {Piron}, {Porter}, {Rain{\`o}}, {Rando},
  {Ransom}, {Ray}, {Razzano}, {Rea}, {Reimer}, {Reimer}, {Reposeur}, {Ripken},
  {Ritz}, {Romani}, {Sadrozinski}, {Sander}, {Saz Parkinson}, {Sgr{\`o}},
  {Siskind}, {Smith}, {Smith}, {Spandre}, {Spinelli}, {Strickman}, {Suson},
  {Takahashi}, {Takahashi}, {Tanaka}, {Thayer}, {Thayer}, {Theureau},
  {Thompson}, {Thorsett}, {Tibaldo}, {Torres}, {Tosti}, {Tramacere},
  {Uchiyama}, {Uehara}, {Usher}, {Vandenbroucke}, {Van Etten}, {Vasileiou},
  {Vilchez}, {Vitale}, {Waite}, {Wang}, {Weltevrede}, {Winer}, {Wood}, {Yang},
  {Ylinen}, \& {Ziegler}}]{LATPWNCat}
{Ackermann}, M. {et~al.} 2011, \apj, 726, 35

\bibitem[{{Atwood} {et~al.}(2009){Atwood}, {Abdo}, {Ackermann}, {Althouse}, \&
  {others}}]{LATinstrumentshort}
{Atwood}, W.~B., {Abdo}, A.~A., {Ackermann}, M., {Althouse}, W., \& {others}.
  2009, \apj, 697, 1071, (LAT)

\bibitem[{{Brazier} {et~al.}(1998){Brazier}, {Reimer}, {Kanbach}, \&
  {Carraminana}}]{Brazier1998}
{Brazier}, K.~T.~S., {Reimer}, O., {Kanbach}, G., \& {Carraminana}, A. 1998,
  \mnras, 295, 819

\bibitem[{{Caraveo} {et~al.}(2010){Caraveo}, {De Luca}, {Marelli}, {Bignami},
  {Ray}, {Saz Parkinson}, \& {Kanbach}}]{Caraveo2010}
{Caraveo}, P.~A., {De Luca}, A., {Marelli}, M., {Bignami}, G.~F., {Ray}, P.~S.,
  {Saz Parkinson}, P.~M., \& {Kanbach}, G. 2010, \apjl, 725, L6

\bibitem[{{Dyks} {et~al.}(2004){Dyks}, {Harding}, \& {Rudak}}]{Dyks2004}
{Dyks}, J., {Harding}, A.~K., \& {Rudak}, B. 2004, \apj, 606, 1125

\bibitem[{{Halpern} {et~al.}(2004){Halpern}, {Gotthelf}, {Camilo}, {Helfand},
  \& {Ransom}}]{Halpern2004}
{Halpern}, J.~P., {Gotthelf}, E.~V., {Camilo}, F., {Helfand}, D.~J., \&
  {Ransom}, S.~M. 2004, \apj, 612, 398

\bibitem[{{Harding} \& {Muslimov}(2011)}]{Harding2011}
{Harding}, A.~K., \& {Muslimov}, A.~G. 2011, \apjl, 726, L10+

\bibitem[{{Hobbs} {et~al.}(2006){Hobbs}, {Edwards}, \&
  {Manchester}}]{Hobbs2006}
{Hobbs}, G., {Edwards}, R., \& {Manchester}, R. 2006, Chinese Journal of
  Astronomy and Astrophysics Supplement, 6, 189

\bibitem[{{Lin} {et~al.}(2010)}]{Lin2010}
{Lin}, C.~C., {et~al.} 2010, \apj

\bibitem[{{Mattox} {et~al.}(1996{\natexlab{a}}){Mattox}, {Bertsch}, {Chiang},
  {Dingus}, {Digel}, {Esposito}, {Fierro}, {Hartman}, {Hunter}, {Kanbach},
  {Kniffen}, {Lin}, {Macomb}, {Mayer-Hasselwander}, {Michelson}, {von
  Montigny}, {Mukherjee}, {Nolan}, {Ramanamurthy}, {Schneid}, {Sreekumar},
  {Thompson}, \& {Willis}}]{mattox1996}
{Mattox}, J.~R. {et~al.} 1996{\natexlab{a}}, \apj, 461, 396

\bibitem[{{Mattox} {et~al.}(1996{\natexlab{b}}){Mattox}, {Koh}, {Lamb},
  {Macomb}, {Prince}, \& {Ray}}]{mattox1996_pulsar}
{Mattox}, J.~R., {Koh}, D.~T., {Lamb}, R.~C., {Macomb}, D.~J., {Prince}, T.~A.,
  \& {Ray}, P.~S. 1996{\natexlab{b}}, \aaps, 120, C95+

\bibitem[{{Muslimov} \& {Harding}(2004)}]{Muslimov04}
{Muslimov}, A.~G., \& {Harding}, A.~K. 2004, \apj, 606, 1143

\bibitem[{{Page} {et~al.}(2009){Page}, {Lattimer}, {Prakash}, \&
  {Steiner}}]{Page2009}
{Page}, D., {Lattimer}, J.~M., {Prakash}, M., \& {Steiner}, A.~W. 2009, \apj,
  707, 1131

\bibitem[{{Pineault} {et~al.}(1993){Pineault}, {Landecker}, {Madore}, \&
  {Gaumont-Guay}}]{Pineault1993}
{Pineault}, S., {Landecker}, T.~L., {Madore}, B., \& {Gaumont-Guay}, S. 1993,
  \aj, 105, 1060

\bibitem[{{Pineault} {et~al.}(1997){Pineault}, {Landecker}, {Swerdlyk}, \&
  {Reich}}]{Pineault1997}
{Pineault}, S., {Landecker}, T.~L., {Swerdlyk}, C.~M., \& {Reich}, W. 1997,
  \aap, 324, 1152

\bibitem[{{Ray} {et~al.}(2011){Ray}, {Kerr}, {Parent}, {Abdo}, {Guillemot},
  {Ransom}, {Rea}, {Wolff}, {Makeev}, {Roberts}, {Camilo}, {Dormody}, {Freire},
  {Grove}, {Gwon}, {Harding}, {Johnston}, {Keith}, {Kramer}, {Michelson},
  {Romani}, {Saz Parkinson}, {Thompson}, {Weltevrede}, {Wood}, \&
  {Ziegler}}]{BlindTiming}
{Ray}, P.~S. {et~al.} 2011, \apjs, 194, 17

\bibitem[{{Romani} \& {Watters}(2010)}]{Romani2010}
{Romani}, R.~W., \& {Watters}, K.~P. 2010, \apj, 714, 810

\bibitem[{{Romani} \& {Yadigaroglu}(1995)}]{Romani1995}
{Romani}, R.~W., \& {Yadigaroglu}, I.-A. 1995, \apj, 438, 314

\bibitem[{{Seward} {et~al.}(1995){Seward}, {Schmidt}, \& {Slane}}]{Seward1995}
{Seward}, F.~D., {Schmidt}, B., \& {Slane}, P. 1995, \apj, 453, 284

\bibitem[{{Slane} {et~al.}(1997){Slane}, {Seward}, {Bandiera}, {Torii}, \&
  {Tsunemi}}]{Slane1997}
{Slane}, P., {Seward}, F.~D., {Bandiera}, R., {Torii}, K., \& {Tsunemi}, H.
  1997, \apj, 485, 221

\bibitem[{{Slane} {et~al.}(2004){Slane}, {Zimmerman}, {Hughes}, {Seward},
  {Gaensler}, \& {Clarke}}]{Slane2004}
{Slane}, P., {Zimmerman}, E.~R., {Hughes}, J.~P., {Seward}, F.~D., {Gaensler},
  B.~M., \& {Clarke}, M.~J. 2004, \apj, 601, 1045

\bibitem[{{Story} {et~al.}(2007){Story}, {Gonthier}, \& {Harding}}]{Story07}
{Story}, S.~A., {Gonthier}, P.~L., \& {Harding}, A.~K. 2007, \apj, 671, 713

\bibitem[{{Verde} {et~al.}(2003){Verde}, {Peiris}, {Spergel}, {Nolta},
  {Bennett}, {Halpern}, {Hinshaw}, {Jarosik}, {Kogut}, {Limon}, {Meyer},
  {Page}, {Tucker}, {Wollack}, \& {Wright}}]{Verde2003}
{Verde}, L. {et~al.} 2003, \apjs, 148, 195

\bibitem[{{Watters} {et~al.}(2009){Watters}, {Romani}, {Weltevrede}, \&
  {Johnston}}]{Watters09}
{Watters}, K.~P., {Romani}, R.~W., {Weltevrede}, P., \& {Johnston}, S. 2009,
  \apj, 695, 1289

\bibitem[{{Yakovlev} {et~al.}(2001){Yakovlev}, {Kaminker}, {Gnedin}, \&
  {Haensel}}]{Yakovlev2001}
{Yakovlev}, D.~G., {Kaminker}, A.~D., {Gnedin}, O.~Y., \& {Haensel}, P. 2001,
  \physrep, 354, 1

\bibitem[{{Ziegler} {et~al.}(2008){Ziegler}, {Baughman}, {Johnson}, \&
  {Atwood}}]{Ziegler2008}
{Ziegler}, M., {Baughman}, B.~M., {Johnson}, R.~P., \& {Atwood}, W.~B. 2008,
  \apj, 680, 620

\end{thebibliography}

\newpage

\begin{center}
\begin{table}
\begin{tabular}{ll}
\hline\hline
\multicolumn{2}{c}{Fit and data-set} \\
\hline
Pulsar name\dotfill & J0007+7303 \\ 
MJD range\dotfill & 54682.7---55415.4 \\ 
Number of TOAs\dotfill & 72 \\
Rms timing residual (ms)\dotfill & 2.3 \\
Reduced $\chi^2$ value \dotfill & 1.39 \\
\hline
\multicolumn{2}{c}{Measured Quantities} \\ 
\hline
Right ascension, $\alpha$\dotfill &  00:07:01.7(2) \\
Declination, $\delta$\dotfill & +73:03:07.4(8)\\
Pulse frequency, $\nu$ (s$^{-1}$)\dotfill & 3.165827392(3) \\ 
First derivative of pulse frequency, $\dot{\nu}$ (s$^{-2}$)\dotfill & $-$3.6120(5)$\times 10^{-12}$ \\ 
Second derivative of pulse frequency, $\ddot{\nu}$ (s$^{-3}$)\dotfill & 4.1(7)$\times 10^{-23}$ \\ 
$\dddot{\nu}$ (s$^-4$)\dotfill & 5.4(9)$\times 10^{-30}$ \\ 
$\Delta\nu$ \dotfill & 0.000001753(2) \\ 
$\Delta\dot{\nu}$ \dotfill & $-$3.5(2)$\times 10^{-15}$ \\ 
\hline
\multicolumn{2}{c}{Set Quantities} \\ 
\hline
Epoch of frequency determination (MJD)\dotfill & 54952 \\ 
Epoch of position determination (MJD)\dotfill & 54952 \\ 
Epoch of dispersion measure determination (MJD)\dotfill & 54952 \\ 
Glitch Epoch \dotfill & 54952.652 \\ 
Phase jump at glitch \dotfill & 0 \\ 
\hline
\multicolumn{2}{c}{Derived Quantities} \\
\hline
$\log_{10}$(Characteristic age, yr) \dotfill & 4.14 \\
$\log_{10}$(Surface magnetic field strength, G) \dotfill & 13.03 \\
\hline
\multicolumn{2}{c}{Assumptions} \\
\hline
Solar system ephemeris model\dotfill & DE405 \\
\hline
\end{tabular}
\caption{Measured and Derived timing parameters of \psr. Figures in parentheses are twice the nominal 1$\sigma$ \textsc{tempo2} uncertainties in the least-significant digits quoted. The time system uses is TDB.}
\label{table:timing}
\end{table}
\end{center}

\begin{table}[ht]
\begin{center}
\begin{tabular}{c c c c c}
  \hline
  $\phi_{min}$ & $\phi_{max}$ &  Photon index & Cutoff energy  &
  Flux ($\geq 100 $ MeV)\\
& & & (GeV) & ($\times$ 10$^{-7}$photons cm$^{-2}$ s$^{-1}$)\\
\hline
\hline
   0.07 & 0.134 & 0.882 $\pm$  2.419 & 0.702 $\pm$  1.854 & 0.393 $\pm$  0.339 \\
 0.134 & 0.182& 1.676 $\pm$  0.269 & 1.529 $\pm$  0.688 & 1.683 $\pm$  0.382 \\
   0.182 & 0.216& 1.791 $\pm$  0.171 & 2.009 $\pm$  0.737 & 4.165 $\pm$  0.581 \\
 0.216 & 0.237& 1.486 $\pm$  0.146 & 1.585 $\pm$  0.381 & 8.759 $\pm$  0.923 \\
 0.237 & 0.253& 1.47 $\pm$  0.107 & 2.613 $\pm$  0.556 & 11.269 $\pm$  1.06 \\
 0.253 & 0.266& 1.362 $\pm$  0.097 & 3.321 $\pm$  0.688 & 12.35 $\pm$  1.1256 \\
 0.266 & 0.28 & 1.479 $\pm$  0.091 & 3.566 $\pm$  0.757 & 13.132 $\pm$  1.148 \\
 0.28 & 0.294 & 1.454 $\pm$  0.099 & 3.606 $\pm$  0.876 & 13.002 $\pm$  1.175 \\
 0.294 & 0.309& 1.374 $\pm$  0.106 & 3.671 $\pm$  0.944 & 11.191 $\pm$  1.0621 \\
 0.309 & 0.324& 1.271 $\pm$  0.096 & 3.414 $\pm$  0.646 & 10.36 $\pm$  0.9783 \\
 0.324 & 0.34  &1.335 $\pm$  0.098 & 3.661 $\pm$  0.748 & 9.723 $\pm$  0.942  \\
 0.34 & 0.357 &1.034 $\pm$  0.111 & 2.606 $\pm$  0.428 & 7.753 $\pm$  0.782 \\
 0.357 & 0.374& 1.285 $\pm$  0.092 & 4.45 $\pm$  0.872 & 8.162 $\pm$  0.811  \\
 0.374 & 0.392& 1.151 $\pm$  0.113 & 3.537 $\pm$  0.777 & 8.073 $\pm$  0.8331 \\
 0.392 & 0.41  &1.247 $\pm$  0.094 & 4.384 $\pm$  0.851 & 7.495 $\pm$  0.747 \\
 0.41 & 0.428 & 1.256 $\pm$  0.106 & 3.524 $\pm$  0.746 & 8.293 $\pm$  0.841 \\
 0.428 & 0.443& 1.264 $\pm$  0.091 & 4.116 $\pm$  0.779 & 9.575 $\pm$  0.91  \\
 0.443 & 0.457 &1.333 $\pm$  0.085 & 4.897 $\pm$  0.966 & 10.91 $\pm$  1.003  \\
 0.457 & 0.471& 1.249 $\pm$  0.085 & 4.755 $\pm$  0.866 & 10.12 $\pm$  0.9369 \\
 0.471 & 0.486& 1.321 $\pm$  0.078 & 5.788 $\pm$  1.124 & 10.768 $\pm$  0.9581 \\
 0.486 & 0.5 &1.111 $\pm$  0.096 & 3.178 $\pm$  0.519 & 10.142 $\pm$  0.95 \\
 0.5 & 0.513 &1.161 $\pm$  0.096 & 3.095 $\pm$  0.523 & 11.402 $\pm$  1.042 \\
 0.513 & 0.528& 1.398 $\pm$  0.081 & 5.504 $\pm$  1.164 & 10.126 $\pm$  0.926 \\
 0.528 & 0.547& 1.483 $\pm$  0.099 & 3.449 $\pm$  0.772 & 9.344 $\pm$  0.892 \\
 0.547 & 0.574& 1.602 $\pm$  0.121 & 3.218 $\pm$  0.921 & 5.652 $\pm$  0.649 \\
 0.574 & 0.623& 1.48 $\pm$  0.263 & 1.502 $\pm$  0.595 & 1.635 $\pm$  0.356 \\
 0.623 & 0.71  &0.762 $\pm$  2.746 & 0.689 $\pm$  1.44 & 0.332 $\pm$  0.405 \\
 \hline
 \hline
\end{tabular}
\end{center}
\caption{Phase interval definitions and corresponding spectral
  parameters obtained from fitting the spectrum with a power law with
  exponential cutoff. The flux in the third column is normalized to
  the width of the phase bin. The systematic uncertainties are in
  agreement with the ones evaluated for the phase-averaged analysis.}
\label{tab:phres}
\end{table}

\begin{table}[ht]
\begin{center}
\begin{tabular}{lcc}
\hline\hline
 & Pre-glitch&Post-glitch \\
\hline
 Date range (MJD)& 54682.68 -- 54952.652 & 54952.652 -- 55412.65 \\
\hline
Photon index &$1.42 \pm 0.04 \pm 0.03$&$1.50 \pm 0.03 \pm 0.04$ \\
Cutoff energy (GeV) &$4.65 \pm 0.39 \pm 0.77 $&$4.74 \pm 0.28 \pm 0.79$\\
Flux ($\geq 100$ MeV) &$3.60 \pm 0.11 \pm 0.27 $&$ 4.09 \pm 0.08 \pm 0.31 $\\
\hline
\end{tabular}
\end{center}
\caption{Phase-averaged spectral parameter and flux for PSR J0007+7303
  for the two epochs around the glitch. Flux is given in $
  10^{-7}$ \pcssp units. First errors are statistical and second ones
  are systematical errors.  }
\label{table:glitch}
\end{table}

\begin{table}[ht]
\begin{center}
\begin{tabular}{c c c c c}
  \hline
Parameter & Outer Gap 1 & Slot Gap 1 & Outer Gap 2 & Slot Gap 2 \\
\hline
\hline
$\alpha$ ($^{\circ}$) & $6^{+1}_{-2}$, $83^{+7}_{-9}$ &
$8^{+4}_{-0}$ & $6^{+1}_{-4}$ & $84$, $8$ \\
$\zeta$ ($^{\circ}$) & $74.5^{+12}_{-3}$, $13.5^{+11}_{-0}$ &
$69.5^{+0}_{-0}$ & $79.5^{+5}_{-5}$ & $11.5^{+2}_{-0}$, $69.5$ \\
$w$ ($r_{\mathrm{ovc}}$) & $0.04^{+0.12}_{-0.03}$,
$0.03^{+0.06}_{-0}$ & $0.0^{+0.01}_{-0}$ & $0.09^{+0.05}_{-0.05}$ &
$0.03^{+0.01}_{-0}$, $0$ \\
$r$ ($R_{\mathrm{lc}}$) & $1.0^{+1.0}_{-0.1}$, $1.0^{+1.0}_{-0}$
& $\geq 1.0$ & $1.1^{+0.9}_{-0.1}$ & $0.7$, $>1.0$ \\
$\Delta \phi$ ($^{\circ}$) & $2^{+0}_{-4}$, $54^{+0}_{-18}$ &
$-4^{+4}_{-0}$ & $10^{+10}_{-10}$ & $186^{+2}_{-0}$, $356$ \\
$\chi^2/27$ & 22.5, 24.5 & 11.3 & 7.0 & 9.3, 21.9 \\
$f_{\Omega}$ & 0.17, 1.10 & 0.42 & 0.15 & 2.0, 0.42 \\
 \hline
 \hline
\end{tabular}
\end{center}
\caption{Best-fit model parameters of the OG and SG geometrical models
to the LAT light curve of PSR J0007+7303, where the background used
for the fitting was 195 counts/bin for columns labeled ``1'' and 246
counts/bin for those labeled ``2''. The asymmetric error bars
give the $3\sigma$ confidence intervals, derived using the scaled
$\chi^2$ as described in \S\ref{sec:LC_modeling}. There were two regions in parameter
space that fell within $3\sigma$ of the best OG1 and SG2 fits; the first set of
values given in the column are the best of the two regions. The quoted
reduced $\chi^2$ and $f_{\Omega}$ correspond to the absolute best fits
in each region. The parameters shown here are used in
Figure~\ref{fig:model_fits} for the simulated light curves and
$\chi^2$ contours.}
\label{table:model_fits}
\end{table}

\acknowledgments
We thank Mallory Roberts and Tyrel Johnson for helpful contributions.

The \textit{Fermi} LAT Collaboration acknowledges generous ongoing
support from a number of agencies and institutes that have supported
both the development and the operation of the LAT as well as
scientific data analysis.  These include the National Aeronautics and
Space Administration and the Department of Energy in the United
States, the Commissariat \`a l'Energie Atomique and the Centre
National de la Recherche Scientifique / Institut National de Physique
Nucl\'eaire et de Physique des Particules in France, the Agenzia
Spaziale Italiana and the Istituto Nazionale di Fisica Nucleare in
Italy, the Ministry of Education, Culture, Sports, Science and
Technology (MEXT), High Energy Accelerator Research Organization (KEK)
and Japan Aerospace Exploration Agency (JAXA) in Japan, and the
K.~A.~Wallenberg Foundation, the Swedish Research Council and the
Swedish National Space Board in Sweden.

Additional support for science analysis during the operations phase is
gratefully acknowledged from the Istituto Nazionale di Astrofisica in
Italy and the Centre National d'\'Etudes Spatiales in France.

This work was performed under contract with the naval research
laboratory, contract N000173-08-2-C004 and was sponsored under a grant
by NASA.

\newpage

\end{document}